\documentclass[11pt]{article}
\usepackage{epsfig}
\begin{document}
\newcommand{\dif}{{\rm d}}
\newcommand{\Tr}{{\rm Tr}}
\newcommand{\ECM}{\em Departament d'Estructura i Constituents de la
Mat\`eria
                  \\ Facultat de F\'\i sica, Universitat de Barcelona \\
                     Diagonal 647, E-08028 Barcelona, Spain\\
                                     and           \\
                                    I. F. A. E.  }
                     
\def\thefootnote{\fnsymbol{footnote}}
\pagestyle{empty}
{\hfill \parbox{6cm}{\begin{center} UB-ECM-PF 98/16\\
                                    August 1998
                     \end{center}}}
\vspace{1.5cm}

\begin{center}
\large{Matching of $U_L(3) \otimes U_R(3)$ and $SU_L(3) \otimes SU_R(3)$  
Chiral Perturbation Theories.}
\end{center}
\vskip .6truein
\centerline {P. Herrera-Sikl\'ody}
\vspace{.3cm}
\begin{center}
\ECM
\end{center}
\vspace{1.5cm}

\centerline{\bf Abstract}
\bigskip
The heavy singlet field is integrated out from the 
$U_L(3) \otimes U_R(3)$ Chiral Perturbation Theory 
and it is shown how its effects 
on the low-energy dynamics 
are reduced to effective vertices for the light mesons. 
The results are matched against  
the standard $SU_L(3) \otimes SU_R(3)$ Chiral 
Perturbation Theory in order to establish the relations
between the coupling constants from both 
theories to one-loop level accuracy. 
\bigskip

\bigskip


\newpage
\pagestyle{plain}

\section{Introduction}

In the low-energy sector, the relevant degrees of freedom in QCD 
are the Goldstone bosons \cite{gold} associated to the spontaneous breaking
of the chiral symmetry: 
$SU_L(3)\otimes SU_R(3) \rightarrow SU_V(3)$. This octet of Goldstone bosons
is identified with the eight lightest pseudoscalar particles: 
the pions, the kaons and the $\eta$; 
their low-energy interactions are well described in terms of the $SU(3)$
Chiral Perturbation Theory or $\chi PT^{[SU(3)]}$ \cite{wein,gl,rev}. 
The model offers good predictions
for energies below a cut-off that is usually set at $M_{\rho}\simeq 770 MeV$.

On the other hand, the classical axial symmetry is also broken, 
but through an anomaly, so it does not generate a ninth
Goldstone boson. Nevertheless, the effects 
of the axial anomaly \cite{adler, belljackiw, adlerbardeen}
are suppressed in $1/N_c$, where
$N_c$ is the number of colors. This means that, 
in the large-$N_c$ limit, one can 
assume a wider scheme containing {\it nine} 
Goldstone bosons (the pseudoscalar octet plus the $\eta'$), 
associated to the spontaneous symmetry breaking $U_L(3)\otimes U_R(3) 
\rightarrow U_V(3)$ \cite{witten1,vene}. 
The corresponding $U(3)$ Chiral Perturbation Theory $\chi PT^{[U(3)]}$
has been described in \cite{vdv,witten,schechter,nosaltres}. 

The relevant point for this paper is that both theories provide 
a good description of the lowest energy range: the {\it smaller} theory is 
the low-energy limit of the {\it bigger} one.
In the low-momenta region, 
the predictions of observables must coincide.
This strong requirement sets the matching conditions
and dictates the relation between the
coupling constants of both theories.

The ninth field $\eta_0$ in $\chi PT^{[U(3)]}$ corresponds 
essentially to the $\eta'$, whose
mass $M_{\eta'}$ is heavier than the typical octet mass. According to
the decoupling theorem \cite{decoup}, if the energy cut-off
is reduced quite below the value of $M_{\eta'}$, the heavy field 
will decouple from the lightest octet fields. 
The dynamics of these remaining fields can then be described 
by a low-energy theory where the heavy degree of freedom does not appear
explicitly: its effects are reduced to effective vertices for the
light fields. At the end of the process, we are left with a theory
where the relevant degrees of freedom are the octet fields. Its
predictions must match those from $\chi PT^{[SU(3)]}$. 

In this particular case --- $\chi PT^{[U(3)]}$ and $\chi PT^{[SU(3)]}$---, 
the matching is enormously simplified by the symmetries 
in both theories. Once the $\eta_0$ field
has been integrated out, the  
resulting theory has the same operator structure than 
$\chi PT^{[SU(3)]}$. This will spare us 
the painful selection and evaluation of observables 
that would be required in general for the matching \cite{lh}.
In this case, the matching can be performed at the effective
Lagrangian level \cite{gl}. 

The singlet field is not an actual physical particle, but a mixing of
the $\pi_0$, the $\eta$ and the heavy $\eta'$ \cite{meta3} instead. 
Strictly speaking, 
the field to be integrated out is $\eta'$, but the resulting theory
would not have the $SU(3)$ symmetry. Furthermore, $M_0^2 \simeq
M_{\eta'}^2$ is a very good approximation, because
the large singlet mass
is a consequence of the anomaly and not of the mixing. 
Therefore, the mixing effects will
be neglected and $\eta'$ will be identified with $\eta_0$. 
On the other hand, the assumption $M_{\eta'} \gg M_{octet}$ 
may not seem numerically
justified, since $M_{\eta'}\sim 2 M_{\eta}$. Both approximations
are nevertheless strongly supported by 
the good results that have been obtained in $\chi PT^{[SU(3)]}$. 

\section{The leading-order Lagrangian and its one-loop effective action}

The nine Goldstone bosons are introduced in the $U(3)$ 
Lagrangian by means of a unitary 
3 $\times$ 3 matrix $\tilde{U}$:
\begin{eqnarray}
\tilde{U} = \exp \left( i \sum_{\alpha =0}^{8} 
\frac{\lambda_{\alpha} \phi _{\alpha}}{f} \right), 
\qquad\mbox{where}\quad \{ \lambda _{\alpha} \}_{\alpha =0, ..., 8}
\quad\mbox{are the $U(3)$ generators.}\quad 
\nonumber
\end{eqnarray}
The 3$\times$3 quark mass matrix 
${\cal M}$ always appears in two combinations of 
$\tilde{\chi}=2 \tilde{B} {\cal M}$:
\begin{eqnarray}
\tilde{M} = \tilde{U}^{\dagger} {\tilde{\chi}}+
{\tilde{\chi}} \tilde{U}, \qquad
\tilde{N} = \tilde{U}^{\dagger} {\tilde{\chi}}-
{\tilde{\chi}} \tilde{U}, \quad \mbox{where} \quad
{\cal M} = \left(\begin{array}{ccc}
m_u & 0 & 0 \\ 0 & m_d & 0 \\ 0 & 0 & m_s  \end{array} \right).
\nonumber
\end{eqnarray} 

The $U(3)$ theory requires two simultaneous expansions: the usual
one, in powers of $p^2/f^2$ and $M^2/f^2$, and the large-$N_c$ expansion, 
in powers of $1/N_c$.
A simple analysis of the particle masses \cite{leut,meta3,kaiser}
leads to the following choice:
$ p^{2}  \sim  m_q  \sim  \frac{1}{N_c}  \sim  \delta$.
The expansion in $\delta$ is the consistent way of working in 
$\chi PT^{[U(3)]}$. Any calculation must be given to a certain 
${\cal O}(\delta)$ accuracy; in each case, the relevant terms in the Lagrangian
will in general mix different orders in momenta. One of the
most interesting features in this way of counting is the fact that
both the leading-order ${\cal O}(\delta)$ and the next-to-leading-order 
${\cal O}(\delta^ 2)$ contribution are tree level.  
Any loop contribution is suppressed by a factor $M^2/f^2 \sim \delta^2$,
so the one-loop diagrams involving leading-order vertices
introduce corrections of ${\cal O}(\delta^3)$ or less.  

The leading-order Lagrangian is ${\cal O}(\delta)$:
\begin{eqnarray}
{\cal L}_{\delta} = \frac{3}{2}\,  v_{02}\,  \eta_0 ^2 + \frac{\tilde{f}^2}
{4} \left(\langle D_{\mu}\tilde{U}^{\dagger} D^{\mu}\tilde{U} \rangle +  
\langle \tilde{M} \rangle \right),
\nonumber
\end{eqnarray}
where $\eta_0=\phi_0$ is the singlet field and brackets stand as usual for 
trace over flavor indices.
$\tilde{B}$, $\tilde{f}$ and $v_{02}$ 
are the free parameters of the theory to be fixed by
experimental data. The tildes are used to distinguish them from
the ones that appear in the $SU(3)$ model.
According to the $N_c$-counting rules, 
$\tilde{f} \sim {\cal O}(N_c^{1/2})$, $\tilde{B} \sim {\cal O}(1) $ and
$v_{02} \sim {\cal O}(N_c ^{-1})$.

The corresponding one-loop effective action can be evaluated 
with the background field method. The fields are decomposed
into a background classical value $\tilde{U}_c$ 
and some quantum fluctuation $\Sigma$:
\begin{eqnarray}
\tilde{U}=\tilde{u}^{\dagger} \Sigma \tilde{u}\, , 
\;\;\; \tilde{U}_c=\tilde{u}^{\dagger}\tilde{u}\,  ,
\;\;\; \Sigma= \exp{\left( i \sum_{\alpha=0}^8 \lambda _{\alpha} 
\Delta_{\alpha} \right)}.
\nonumber
\end{eqnarray}
(Notice that the $\Delta_0$ fluctuations factorise in a natural way
from the rest of the fields because $\lambda_0$ commutes with everything).
The action is then expanded in powers of the fluctuations 
up to quadratic terms and the path integral is performed over
all possible configurations of these fluctuations.
At the end, the effective action will include
the bare Lagrangian 
${\cal L}_{\delta}$ itself, its one-loop corrections, that are
${\cal O}(\delta^3$),
and the appropriate  ${\cal O}(\delta^2$) and ${\cal O}(\delta^3$) counterterms
required to cancel the divergences. 
\begin{eqnarray}
\Gamma_{eff}^{one-loop} [\tilde{U}_c] = 
  \int \dif^4 x \left( {\cal L}^r_{\delta}(\tilde{U}_c)+
{\cal L}^r_{\delta^2}(\tilde{U}_c) +
{\cal L}^r_{\delta^3}(\tilde{U}_c)\right)+
\;\mbox{finite one-loop corrections.}
\nonumber
\end{eqnarray}
${\cal L}^r_{\delta^n}$ stands for the renormalized Lagrangian of
order $\delta^n$. 
Schematically, these Lagrangians
are built with the operators associated to the following list of
coupling constants (see \cite{nosaltres} for the complete list of operators):
 \begin{eqnarray}
{\cal L}_{\delta^2} &:& v_{31}, L_i(0) \;\;\;\; i = 1, 2, 3, 5, 8, 9, 10, 
11, 12, 13, 14, 15, 16. \nonumber \\
{\cal L}_{\delta^3} &:& v_{04}, v_{12}, v_{22}, L_i(0) \;\;\;\; i = 4, 
6, 7, 18, 19, 20.
\nonumber
\end{eqnarray}

For sufficiently low energies, the heavy field $\eta_0$ 
appears only as a fluctuation and its background value is zero.
To ${\cal O}(\Delta^2)$, the ${\cal O}(\delta$) Lagrangian  
has then the following structure:
\begin{eqnarray}
{\cal L} (\tilde{U})\big|_{\eta_0=0} \sim {\cal L}(U_c) 
+ J_0 \,  \Delta_0 - \frac{1}{2} \, \Delta_0\,  D_{00} \, 
\Delta_0 - \frac{1}{2} \, \Delta_a\,  D_{ab}\, \Delta_b -
\Delta_a\,  D_{a0}\, \Delta_0,  
\label{exp1}
\end{eqnarray}
where repeated indices are to be summed over a, b = 1, ..., 8, and
$J_0$, $D_{00}$, $D_{0a}$ and $D_{ab}$ are functions of $U_c=
\tilde{U}_c \big|_{\eta_0=0}$. 
The 9$\times$9 matrix $D$ can be written as:
\begin{eqnarray}
D_{\alpha \beta}= (\tilde{d}_{\mu}\tilde{d}^{\mu} + \tilde{\sigma} 
)_{\alpha \beta}\, \bigg|_{\eta_0=0}, 
\;\;  \alpha,\beta = 0, ..., 8. 
\label{D}
\end{eqnarray}
The complete expressions for $\tilde{\sigma}$ and $\tilde{d}_{\mu}$ 
have been given in \cite{nosaltres}
\footnote{Notice that all the tilded symbols that appear in the present paper
are untilded in \cite{nosaltres}, but the double notation is needed now
to distinguish the cases $\eta_0=0$ and $\eta_0\neq 0$.}. 
The covariant derivative $\tilde{d}_{\mu}$
includes a connection: $
(\tilde{d}_{\mu}\Delta)^{\alpha} = \partial_{\mu} \Delta^{\alpha}+
\omega^{\alpha \beta}_{\mu} \Delta^{\beta}$, 
although we will not be concerned about its particular form,
because the ${\cal O}(\delta)$ Lagrangian gives $\omega_{0\alpha}=0$ 
$\forall \alpha$.
Similarly, the rest is reduced to $\tilde{\sigma}\big|_{\eta_0=0}=
\sigma$. At the end, the relevant objects are:
\begin{eqnarray}
\begin{array}{cc}
D_{00}=\partial_{\mu}\partial^{\mu} + \sigma_{00},\qquad & \qquad  
\sigma_{00} = \frac{1}{6} \,\langle \tilde{M} \rangle -3 \, v_{02} = 
M_0^2 + \hat{\sigma}, \\
D_{a0} = D_{0a} \; = \; \sigma_{a0} \; = \; \frac{1}{2 \sqrt{6}}
 \, \langle \lambda_a \tilde{M} \rangle,  \quad  a \neq 0, &
J_0 = i \, \frac{\tilde{f}}{2 \sqrt{6}}  \, \langle \tilde{N} \rangle. 
\end{array}
\end{eqnarray}

Before the integration, it is convenient to 
diagonalize the quadratic part in (\ref{exp1}) by
means of a change of variables:
\begin{eqnarray}
\varphi_a = \Delta_a + (D^{-1} )_{ab} D_{b0}\,  \Delta_0 ,\qquad
\varphi_0 = \Delta_0.\nonumber
\end{eqnarray} 
The resulting expression exhibits a perfect quadratic structure:
\begin{eqnarray}
{\cal L} (\tilde{U}) \big|_{\eta_0=0} \simeq {\cal L}(U_c) + 
J_0\,  \varphi_0 -  \frac{1}{2} \, \varphi_0 \left(
D_{00} - D_{0a}(D^{-1} )_{ab}D_{b0} \right) \varphi_0 - 
\frac{1}{2}\, \varphi_a\,  D_{ab}\, \varphi_b. 
\label{expd}
\end{eqnarray}

The effective action is given by integrating over all configurations for
the fluctuations ($\varphi_0$ and $\varphi_a$'s):
\begin{eqnarray}
e^{\, i \Gamma_{eff}^{[U(3)]} [U]} = \int [\dif \varphi_0]\prod_{a=1}^8 
[\dif \varphi_a] \, e^{ i 
{\int \dif ^4 x \, {\cal L}(\tilde{U})}} \, {\bigg |}_{\eta_0 = 0}\, .
\nonumber
\end{eqnarray}
In order to get a more friendly notation, the subscript $c$ has been
dropped. In what follows, every $U$ or $\tilde{U}$ is to be understood
as made of classical fields whose value is set to the background value. 

A straightforward Gaussian integration leads to
\begin{eqnarray}
\Gamma_{eff} [U] &=& \frac{1}{2} \int \dif ^4x \, J_0 \, D_{00}^{-1}\, J_0 + 
\frac{i}{2} \, \Tr\,  \ln\,D_{00} -
\frac{i}{2}\,  \Tr \left(D_{00}^{-1}D_{0a}(D^{-1} )_{ab}D_{b0} \right) 
\nonumber \\
&+& \int \dif ^4 x\, {\cal L}_{U(3)} (U)  +\frac{i}{2}\, \Tr\,\ln D_{ab} + 
{\cal O}(\delta^3).
\label{gamma}
\end{eqnarray}
The last term (where the sub-indices have been kept to recall that it
is an 8$\times$8 matrix) contains all the one-loop diagrams 
with particles from the octet circulating in the internal lines.
The three terms in the first line are due to diagrams containing one and two 
{\it heavy} internal lines and will be analyzed below.
All these one-loop contributions contain divergences that are absorbed
by the appropriate counterterms, so we will be
dealing with one-loop renormalized coupling constants.

\vspace{5pt}
The first term in the right-hand side of (\ref{gamma}) corresponds to
a tree graph with light external lines and a heavy internal 
$\eta_0$ propagator that is seen as an $SU(3)$ effective vertex in the
low-energy theory (fig. 1 a). This term was already
analyzed in \cite{gl,drp}. 
The operator $D_{00}$ can be split into two
pieces:
\begin{eqnarray}
D_{00} = D_o + \hat{\sigma}. 
\label{sigmahat}
\end{eqnarray}
$D_o$ is the free operator of a scalar field with mass $M_0$:
$ D_o= \partial_{\mu} \partial^{\mu} + M_0 ^2$, and 
$\hat{\sigma}= \sigma_{00}- M_0^2$.  $\hat{\sigma}$  contains 
vertices with two or more light fields. Its inclusion 
in the following calculation would only contribute to ${\cal O}(p^6)$ vertices,
so it will not be considered.

If the cut-off is small compared to $M_0^2$, or in the limit of
large distances, the interaction can be assumed to be local
and the following integral can be approximated
by a delta function:
\begin{eqnarray}
D_{00}(x) \simeq \int \frac{\dif ^4 p}{(2\pi)^4} 
\, e^{- i p \cdot x} \frac{1}{M_0^2-p^2} \simeq 
\frac{1}{M_0^2} \, \delta^4 (x).
\nonumber
\end{eqnarray}
Thus the contribution due to this term is:
\begin{eqnarray}
\Gamma_{tree} [U] = - \frac{\tilde{f}^2}{48 \, M_0^2}\, 
\int \dif ^4x \, \langle \tilde{N} 
\rangle ^2.
\label{tree}
\end{eqnarray}

The other pieces originate in one-loop graphs.
The identity (\ref{sigmahat}) can be used to expand the trace of the 
logarithm in (\ref{gamma}):
\begin{eqnarray}
\frac{i}{2}\,  \Tr \, \ln D_{00} = 
\frac{i}{2} \, \Tr \, \ln (D_o + \hat{\sigma}) 
\simeq  \frac{i}{2}\, \Tr\,  \ln D_o + \frac{i}{2}\, \Tr \left(D_o^{-1}
 \hat{\sigma}\right) - \frac{i}{4}\,  \Tr \left(D_o^{-1} \hat{\sigma}
 D_o^{-1} \hat{\sigma}\right) + ...
\label{explog}
\end{eqnarray}
The first term in (\ref{explog}) is a constant that can be ignored.
The second and third terms correspond to one-loop diagrams with one and two  
internal $\eta_0$ lines, respectively (fig.1 b and c).   
 
The $\eta_0$ tadpole term, that we shall call  
$\Gamma_{\eta_0}$, gives:
\begin{eqnarray}
\Gamma_{\eta_0}[U] &=& \frac{i}{2}\, \Tr \left(D_o^{-1}\hat{\sigma}\right)
\;= \; \frac{i}{2} \, \Delta_0 (0) \int \dif ^4 x\,  
\hat{\sigma}(x) \nonumber \\
&=&  -\frac{1}{6}\, (M_0^2 \lambda_{\epsilon} + 
\frac{M_0^2}{32 \pi^2}\, \ln \frac{M_0^2}{\mu^2}) \int \dif ^4 x 
\, \left(\langle \tilde{M} \rangle-18 \,v_{02}-6\, M_0^2\right) ,
\label{etao} 
\end{eqnarray}
where $\lambda_{\epsilon}$ is a divergent term and is given in the
appendix (\ref{lam}). This and the other divergent pieces that
will come up in (\ref{2etao}) and (\ref{etaopi}) are just part of the 
one-loop renormalization of the leading-order U(3) theory.

$\Gamma_{\eta_0\eta_0}$ contains all the diagrams
with two internal $\eta_0$.  For
the sake of clearness, the details of the calculation 
have been relegated to the appendix. The resulting contribution is:
\begin{eqnarray}
\Gamma_{\eta_0\eta_0}[U] &=& -\frac{i}{4} \, \Tr \left( D_o^{-1}\hat{\sigma}
D_o^{-1}\hat{\sigma} \right) 
\; = \; -\frac{1}{2} \, (k_{00}+\lambda_{\epsilon}) \int \dif ^4 x \, 
\hat{\sigma}(x)^2 \nonumber \\
&=& -\frac{1}{72} \, (k_{00}+\lambda_{\epsilon}) \int \dif ^4 x \, \left(\langle 
\tilde{M} \rangle  -18 \, v_{02}-6 \, M_0^2 \right) ^2.
\label{2etao}
\end{eqnarray}
$k_{00}$ is given in (\ref{kas}).
Obviously, the constant terms in (\ref{etao}) and (\ref{2etao}) 
can be dropped out.

Finally, the third term in (\ref{gamma}) corresponds to 
one-loop diagrams with one internal
$\eta_0$ and one internal $\pi$
(unless otherwise stated, {\it pion} is used in a generic sense, 
meaning any particle from the octet). In the $U(3)$ theory, 
the pions can 
take higher values of momentum and these modes must be integrated
out, too. This integration can be understood in two different steps:
the integration of heavy $\eta_0$'s and of {\it high} momenta pions 
yields an $SU(3)$ {\it bare}
vertex with pion external legs and one internal pion line with low
momenta. The integration of these remaining low-momentum modes 
gives the $SU(3)$ tadpole renormalization of this new vertex (fig. 1 d). 

The reader should again refer to the appendix for the details of the
calculation. The result in this case is (P labels the octet mesons):
\begin{eqnarray}
\Gamma_{\eta_0\pi}[U] &=& -\frac{i}{2} \, \Tr \left( (D^{-1})_{ab} \sigma_{a0} 
D^{-1}_{00}\sigma_{0b}\right) 
= - \sum_{P}(\lambda_{\epsilon}+ k_{0P}) \int \dif ^4 x \, 
\sigma_{0P}(x)^2 \nonumber \\
&=& -\frac{1}{12} \left( \lambda_{\epsilon}+\frac{1}{32\pi^2}\ln\, \frac{M_0^2}
{\mu^2} \right) \int \dif ^4 x \, \left(
\langle \tilde{M}^2 \rangle - \frac{1}{3}\langle \tilde{M}
 \rangle^2 \right) \nonumber \\
&+& \frac{1}{24\cdot 32 \pi^2} \sum_{P} a_P
\int \dif ^4 x \, \langle \tilde{M} \lambda_P \rangle^2,  
\label{etaopi}
\end{eqnarray}
where $k_{0P}$ is given in (\ref{kas}) and $a_P$ is defined as:
\begin{eqnarray}
a_P= \frac{M_P^2}{M_0^2-M_P^2} \,\ln\frac{M_P^2}{M_0^2}.
\nonumber
\end{eqnarray}
The last term in (\ref{etaopi}) reflects 
the explicit breaking of the $U(3)$ symmetry.
This contribution could be neglected if all quark masses were small enough: 
the corrections depend on the ratio $M_P^2/M_0^2$. It will be taken into
account, however, because
$M_K^2$ and $M_{\eta}^2$ are not that small when compared to $M_0^2$. 
This happens because $m_s \gg m_u$, $m_d$.
For simplicity, we shall assume that $m_u, m_d 
= 0$ but $m_s \neq 0$. In this limit, $M_{\pi}=a_{\pi}=0$ and
$\chi$ turns out to be a very simple matrix; as a consequence,
\begin{eqnarray}
\langle \tilde{M} \lambda_i \rangle = 0 \quad \mbox{for} \; i=1,\, 2,\, 3,
 \qquad \qquad
\lambda_8 = \frac{1}{\sqrt{3}} \left( I -\frac{3}{2\,B\,m_s}\,
\chi \right),
\\ \nonumber 
\sum_{i=4}^7 \langle \tilde{M} \lambda_i \rangle^2 = 
2 \langle \tilde{M}^2 \rangle - \frac{2}{3}\,  \langle \tilde{M} \rangle ^2 -
\langle \tilde{M} \lambda_8 \rangle^2.\qquad \qquad 
\nonumber
\end{eqnarray}
Recall that the $\eta-\eta'$ mixing effects are neglected, 
so $\lambda_{\eta}\simeq \lambda_8$.
At the end, one can write:
\begin{eqnarray}
\sum_{P} a_P \, \langle \tilde{M} \lambda_P \rangle^2 \simeq 2 \, a_K \, 
\langle \tilde{M}^2 \rangle + (\frac{a_{\eta}}{3}- a_K) \, 
\langle \tilde{M} \rangle^2 +
{\cal O}(\frac{M_{\pi}^ 2}{M_0^2}) + {\cal O}(p^6).
\nonumber
\end{eqnarray}
\vspace{5pt}

\begin{figure}
\epsfig{file=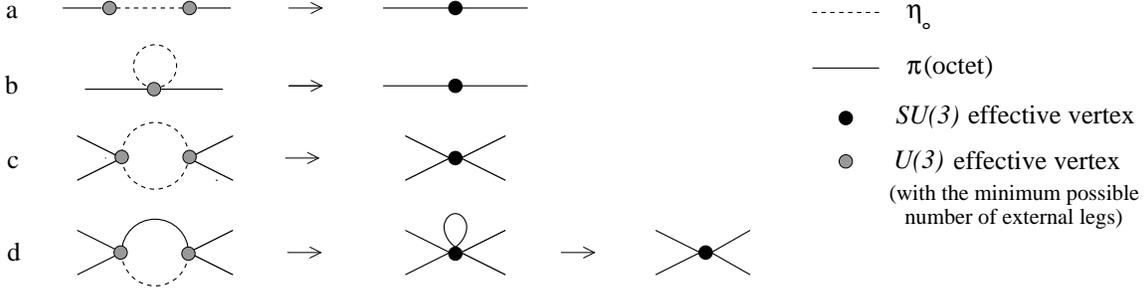,width=150mm}
\caption{The effects of heavy internal modes reduce to effective vertices in
the $SU(3)$ theory.}
\end{figure}

\section{Matching}

The low-energy one-loop effective action (\ref{gamma}) is given by the sum  
of (\ref{tree}), (\ref{etao}), (\ref{2etao}) and (\ref{etaopi}):
\begin{eqnarray}
\Gamma^{U(3)}_{eff} [U] &=& \int \dif ^4 x\, {\cal L}^r_{U(3)} (U) + 
\; \mbox{finite corrections involving one loop of pions} 
- \frac{\tilde{f}^2}{48 \, M_0^2}\, \langle \tilde{N} \rangle ^2 \nonumber \\ 
&+& \frac{M_0^2}{6 \cdot 32 \pi^2 \tilde{f}^2}\, \ln \frac{M_0^2}{\mu^2}
\int \dif ^4 x \, \langle \tilde{M} \rangle 
-\frac{1}{72}\, k_{00} \int \dif ^4 x \, \left( \langle \tilde{M} \rangle ^2 
-12\, (M_0^2 +3 v_{02}) \, \langle \tilde{M} \rangle \right)
\nonumber \\
&-& \frac{1}{12\cdot 32 \pi^2} \ln\, \frac{M_0^2}{\mu^2} 
\int \dif ^4 x \, \left(\langle \tilde{M}^2 \rangle - 
\frac{\langle \tilde{M} \rangle^2}{3} \right) 
+ \frac{1}{24\cdot 32 \pi^2} \sum_{P} a_P
\int \dif ^4 x \, \langle \tilde{M} \lambda_P \rangle ^2.
\label{total}
\end{eqnarray}

This effective action describes
the {\it same system} as the $SU(3)$ one-loop effective action \cite{gl}: 
\begin{eqnarray}
\Gamma^{SU(3)}_{eff} [U] &=& \int \dif ^4 x\, {\cal L}_{SU(3)} (U)  +
\frac{i}{2}\, \Tr\,\ln D_{ab} \nonumber \\ 
&=& \int \dif ^4 x\, {\cal L}^r_{SU(3)} (U) + \; \mbox{finite 
corrections involving one loop of pions}.
\label{gammasu3}
\end{eqnarray}

The matching conditions require physical quantities to 
give identical results in both cases. In a general case, one would
be forced to compare the observables that stem from both theories.
At tree-level, for instance, both theories give a prediction for
the mass of the pion, and they must be equal: $ B(m_u+m_d)= 
\tilde{B}(m_u+m_d)$. This implies that $\tilde{B}= B+{\cal O}(\delta^2)$. 

In this case, however, one needs not go all the way down to  
observables. The operator structure of 
both effective actions is identical, allowing for a much easier procedure: 
\begin{eqnarray}
\Gamma^{SU(3)}_{eff} [U]= \Gamma^{U(3)}_{eff} [U].
\label{igual}
\end{eqnarray}
One can safely replace $\tilde{B}$ by $B$ everywhere in 
(\ref{total}) except
in the original ${\cal O}(\delta)$ 
Lagrangian, because $\tilde{B}= B+{\cal O}(\delta^2)$. 
By doing this, all the {\it corrections involving
one loop of pions} in (\ref{gammasu3}) and (\ref{total}) become identical
and cancel out in (\ref{igual}). Both theories must indeed present 
identical behaviors
in the IR region. In particular, the IR non-analyticities that occur in 
these pion-loop terms in
the chiral limit are exactly the same and the matching calculation is
IR finite \cite{georgi}.
 
All the ${\cal O}(p^4)$ terms ---except for
the last one--- can then be easily
written in terms of the usual $SU(3)$ operators that include the
external source $\chi=2 B {\cal M}$:
\begin{eqnarray}
\langle M^2\rangle =  O_8 +2 \, O_{12}, \qquad
\langle M\rangle^2 =  O_6, \qquad
\langle N\rangle^2 =  O_7,
\nonumber
\end{eqnarray}
where: 
\begin{equation}
\begin{array}{ll}
O_6 =  \langle U^{\dag}\chi+\chi^{\dag} U \rangle^2 , &
O_7 = \langle U^{\dag}\chi-\chi^{\dag} U \rangle^2,\\
O_8 = \langle U^{\dag}\chi U^{\dag}\chi+
\chi^{\dag} U \chi^{\dag} U \rangle,\qquad &
O_{12} = \langle \chi^{\dag} \chi \rangle .
\end{array}
\nonumber
\end{equation}

Finally, by directly comparing the structures and 
identifying the factors preceding  each operator on both sides of
(\ref{igual}), one obtains the following relations between the
renormalized $SU(3)$ coupling constants (plain) and the $U(3)$ ones
(tilded), in terms of physical quantities
\footnote{$M_0^2 \simeq M_{\eta'}^2$ and $-3 \, v_{02}= M_{\eta'}^2+
M_{\eta}^2-2\, M_K^2 + {\cal{O}}(\delta^2)$}:
\begin{eqnarray}
f &=&\tilde{f}, \nonumber \\
B &=&  \tilde{B}(\mu)+ \frac{\tilde{B}(\mu)}{48 \pi^2 f_{\pi}^2}
\left( M_{\eta'}^2 \, \ln \frac{M_{\eta'}^2}{\mu^2}  + 
(M_{\eta}^2-2\, M_K^2) \, 
(\ln \frac{M_{\eta'}^2}{\mu^2}+1)
\right), \nonumber \\
L_6^r(\mu) &=& \tilde{L}_6^r(\mu) + \frac{1}{72 \cdot 32 \pi^2} 
\left( \ln \frac{M_{\eta'}^2}{\mu^2} + \, a_{\eta} - 3 \, a_K -1
\right)+ {\cal O}(\frac{M_{\pi}^ 2}{M_{\eta'}^2}), 
\nonumber \\
L_7^r(\mu) &=& \tilde{L}_7^r(\mu) - \frac{f_{\pi}^2}{48\, M_{\eta'}^2}, 
\nonumber \\
L_8^r(\mu) &=& \tilde{L}_8^r(\mu)- \frac{1}{12\cdot 32 \pi^2}\left( 
\ln \frac{M_{\eta'}^2}{\mu^2}- a_K
\right)+ {\cal O}(\frac{M_{\pi}^ 2}{M_{\eta'}^2}), \nonumber \\
H_2^r(\mu) &=& \tilde{L}_{12}^r(\mu) - \frac{1}{6\cdot 32 \pi^2} \left(
\ln \frac{M_{\eta'}^2}{\mu^2}-a_K
\right)+ {\cal O}(\frac{M_{\pi}^ 2}{M_{\eta'}^2}), \nonumber \\
L_i^r(\mu) &=&  \tilde{L}_i^r(\mu),
\qquad \mbox{ for i = 1, 2, 3, 4, 5, 9, 10}, \nonumber \\
H_1^r(\mu) &=& \tilde{L}_{11}^r(\mu) .
\label{relations}
\end{eqnarray}
One can check that the dependence on the renormalization
parameter $\mu$ is the same on both sides of the equalities. 
This parameter has been introduced to deal with the UV divergences in both
theories; a convenient choice of its value will avoid the growth of large logs 
and the breakdown of perturbation theory. 
The typical scale used in $\chi PT^{[SU(3)]}$, $\mu \simeq M_{\rho}$, 
will do the job.  

\section{Concluding remarks}

It is worth emphasizing the running of $\tilde{B}$, because this is a
special characteristic of $\chi PT^{[U(3)]}$:
\begin{eqnarray}
\mu \frac{\partial \tilde{B}^r}{\partial \mu} &=& \tilde{B}^r\frac{v_{02}}
{16\,\pi^ 2 f^2}\, .
\nonumber
\end{eqnarray}
The correction to $\tilde{B}$ exhibits the typical 
$M_{\eta'}^2$ correction to the light
masses that arises from the integration of a heavy scalar field. This results 
in a paradox ---the so-called naturalness problem--- 
when one tries to push the calculation to the limit 
$M_{\eta'} \rightarrow \infty$. The contradiction is easily solved in this 
case: 
if $\eta'$ were {\it very} heavy, 
the nonet theory that we started from would be wrong. 

As expected from the Appelquist-Carrazone theorem, all the effects
from the integrated heavy particle are either suppressed in powers of
$M_{\eta'}^2$ and/or can be re-absorbed in the coupling constants of the 
lower theory.
The correction to $L_7$, for instance, 
originates in the momentum expansion of a perfectly analytical tree graph. 
In contrast, loop graphs contributions incorporate 
non-analytical $\ln M_{\eta'}$ terms --- that could be never obtained through
a Taylor expansion.
 
The value of the coupling constants in the $SU(3)$ theory are
relatively well known, so this work
offers a first estimate of 
the unknown $U(3)$ parameters.
A numerical check can be done in the case of $L_8$. 
At $\mu=M_{\rho}$, $L_8(M_{\rho})= (0.9 \pm 0.3) \cdot  10^{-3}$ 
and the value predicted in 
(\ref{relations}) 
is $\tilde{L_8}(M_{\rho})= (1.2 \pm 0.3) \cdot 10^{-3}$. 
This is too small compared to
the value found in \cite{meta3}, where $\tilde{L}_8$ was estimated to be
$(1.3-1.6) \cdot 10^{-3}$, but the correction goes in the
right direction.
 
$L_7$ has always been related to $\eta'$.
This has produced some confusion on the $N_c$-power counting 
of this parameter \cite{gl,drp}.
The problem disappears by noticing that the  
$1/N_c$ expansion {\sl must} be implemented in
the $U(3)$ context: the $N_c \rightarrow \infty$ limit has no meaning
in the $SU(3)$ theory, because the very first consequence of assuming
the large-$N_c$ limit is that there are {\sl nine} Goldstone bosons instead of
eight. One might however wish to keep track of the $N_c$ counting 
for each $SU(3)$ coupling because it justifies why some of them
are smaller than the rest and thus negligible. It can be seen in
(\ref{relations}) that all corrections but the correction for $\tilde{L}_7$
are either ${\cal O}(1)$ or suppressed in
$\delta^2$, so the $N_c$ counting for the $SU(3)$ couplings stays the same 
as in $U(3)$, except
for one case: $\tilde{L}_7$ is ${\cal O}(1)$, but its correction is
${\cal O}(N_c^2)$, so $L_7$ ends up being ${\cal O}(N_c^2)$. 
The numerical value of this correction is $-0.2 \cdot 10^{-3}$,
which has indeed the same order of magnitude of the present value of 
$L_7 = (-0.4\pm 0.2) \cdot 10^{-3}$. 

\section{Acknowledgments}

The author is grateful to J. I. Latorre, J. Taron for 
their suggestions and critical reading of the manuscript. 
Discussions with them 
and with P. Pascual were very helpful. She also acknowledges a Grant
from the {\sl Generalitat de Catalunya}. This work has been financially
supported by CICYT, contract AEN95-0590, and by CIRIT, contract GRQ93-1047.  

\section{Appendix}

The tadpole term (\ref{etao}) reduces to $\Delta_P(0)$,
the Feynman propagator of a scalar particle of mass $M_P$ in
$z=0$.
\begin{eqnarray}
i \Delta_P(0) = -2\, M_P^2 \, \lambda_{\epsilon}-\frac{M_P^2}{16 \pi^2}
\, \ln\, \frac{M_P^2}{\mu^2}, \quad \mbox{where} \quad
\lambda_{\epsilon}= \frac{\mu^{2 \epsilon}}{2\, (4 \pi)^2} \left( 
\frac{1}{\epsilon} + \gamma - \ln 4\pi -1 \right) .
\label{lam} 
\end{eqnarray}

The traces in (\ref{2etao}) and (\ref{etaopi}) involve a particular
kind of integral and can be written in terms of
a function $J_{P0}(z)=-i \Delta_0(z)\Delta_P(z)$:
\begin{eqnarray}
-\frac{i}{2}\, \Tr \left( D_{ab}^{-1}\sigma_{a0}D_o^{-1}\sigma_{0b} \right) &=&
\frac{1}{2}\int \dif ^4 x \, \dif ^4 y \, J_{0P}(x-y) \, \sigma_{0P}(x)\, 
\sigma_{0P}(y) ,
\nonumber \\
-\frac{i}{4}\, \Tr \left( D_o^{-1}\hat{\sigma}D_o^{-1}\hat{\sigma} \right) &=&
\frac{1}{4} \int \dif ^4 x \, \dif ^4 y \, J_{00}(x-y) \, \hat{\sigma}(x)\, 
\hat{\sigma}(y) ,
\label{J}
\end{eqnarray}
In momentum space and using dimensional regularization, $D=4+2\epsilon$,  
\begin{eqnarray}
J_{0P}(s) &=& \int \dif^4 z \; e^{ipz} \; J_{0P}(z) 
= -i \int \frac{\dif ^D q}{(2 \pi)^D} \;
\frac{1}{M_0^2-q^2+i\varepsilon} \; \frac{1}{M_P^2-(p-q)^2+i\varepsilon} 
\nonumber \\ 
&=& -2 \,\lambda_{\epsilon} -2 \, k_{0P} + \bar{J}_{0P}(s),  
\nonumber
\end{eqnarray}
where $s=p^2$ is the external momentum and
\begin{eqnarray}
k_{0P} =\frac{1}{32 \pi^2} \left(\ln \frac{M_0^2}{\mu^2} +
\frac{M_P^2}{M_P^2-M_0^2}\, \ln \, \frac{M_P^2}{M_0^2}\right) ,
\qquad
k_{00} =\frac{1}{32 \pi^2} \left(\ln \frac{M_0^2}{\mu^2} + 1 \right) .
\label{kas}
\end{eqnarray}
$\bar{J}_{0P}(s)$ is some function of s, $M_0^2$ and $M_P^2$,
but we shall omit it, because it does not contribute to the low-energy limit:
\begin{eqnarray}
J_{0P}(s) \simeq J_{0P}(0) + {\cal O}(\frac{p^2}{M_0^2}) = 
-2 \,\lambda_{\epsilon} - 2 \,k_{0P}
 + {\cal O}(\frac{p^2}{M_0^2}).
\nonumber
\end{eqnarray}

In this limit, the integrals (\ref{J}) reduce to a very simple form:
\begin{eqnarray}
-\frac{i}{2}\, \Tr \left( D_{ab}^{-1}\sigma_{a0}D_o^{-1}\sigma_{0b} \right) &=&
-\sum_P\left(  k_{0P}+\lambda_{\epsilon}   \right) \int \dif ^4 x \; 
\sigma_{0P}(x)^2, \nonumber \\
-\frac{i}{4}\, \Tr \left( D_o^{-1}\hat{\sigma}D_o^{-1}\hat{\sigma} \right) &=&
-\frac{1}{2}\left(   k_{00}+\lambda_{\epsilon}   \right) \int \dif ^4 x 
\; \hat{\sigma}(x)^2.
\nonumber
\end{eqnarray}

\end{document}